\documentstyle[epsfig,subfigure,amssymb]{europhys}

\def\And{{\rm and\ }}

\newif\ifboo \boofalse

\def\Review#1{\boofalse{\it #1},}
\def\Name#1{{\sc #1},}
\def\Vol#1{\ifboo Vol. {\bf #1}\else{\bf #1}\fi}
\def\Year#1{\ifboo #1\else(#1)\fi}
\def\Book#1{\bootrue{\it #1},}
\def\Page#1{\ifboo {\rm p. #1}\else{\rm #1}\fi}

\begin{document}
\euro{vv}{nn}{ppp--ppp}{1999} 
\Date{1999}
\shorttitle{G. CUNIBERTI {\it et al.}: FREQUENCY SCALING OF 
      PHOTO--INDUCED TUNNELING}

\title{Frequency scaling of photo--induced tunneling.}

\author{G. Cuniberti\inst{1}$^{,}$\inst{2},
        A. Fechner\inst{2}$^{,}$\inst{3},
        M. Sassetti\inst{2}, and
        B.  Kramer\inst{3}
       }

\institute{\inst{1} Max-Planck-Institut f{\"u}r Physik komplexer Systeme\\
                    N{\"o}thnitzerstra{\ss}e 38, D-01187 Dresden, Germany\\
           \inst{2} Dipartimento di Fisica, INFM,
                    Universit{\`a} di Genova\\
                    via Dodecaneso 33, I-16146 Genova, Italy\\
           \inst{3} I. Institut f\"ur Theoretische Physik, 
                    Universit{\"a}t Hamburg\\
                    Jungiusstra{\ss}e 9, D-20355 Hamburg, Germany
          }

\rec{dd Month 1999}{dd Month 1999 in final form}

\pacs{
\Pacs{71}{10Pm}{Fermions in reduced dimensions}
\Pacs{72}{20Ht}{High--field and nonlinear effects}
\Pacs{42}{65Ky}{Harmonic generation, frequency conversion}
      }
\maketitle

\begin{abstract}
  The DC current-voltage characteristics induced by a driving electric
  field with frequency $\Omega$ of a one dimensional electron channel
  with a tunnel barrier is calculated. Electron-electron interaction
  of finite-range is taken into account. For intermediate interaction
  strengths, the non-linear differential conductance shows cusp-like
  minima at bias voltages $m\hbar\Omega/e$ ($m$ integer) that are a
  consequence of the finite non-zero range of the interaction but are
  independent of the shape of the driving electric field. However, the
  frequency-scaling of the photo-induced current shows a cross-over
  between $\Omega ^{-1}$ and $\Omega ^{-2}$, and depends on the
  spatial shape of the driving field and the range of the interaction.

\end{abstract}

The influence of a mono-chromatic driving field on the transport
through tunnel barriers has become an extensively studied subject.  A
paradigmatic result has been obtained more than three decades ago by
Tien and Gordon for a superconducting tunnel junction \cite{tg63}. Similar
approaches have been used to study photo-assisted effects in the
tunneling through quantum point contacts \cite{hu93} and driven
quantum wells \cite{w96}. Basically, the frequency of the driving
field has been shown to produce sidebands due to the non-linearity of
the current-voltage characteristics. In all of these works,
interaction between the electrons has been neglected. Recently, the
influence of electron-electron interaction on non-linear tunneling
transport has been studied in the regions of charging, interaction
\cite{mw92,bs94,n96} and strong correlations
\cite{kf92,swk96,sk96,fsk99}. In mean field approximation,
renormalisation of the driving field has been studied \cite{pb98}.
Also experimentally, in view of the realization of the quantum dot
current turnstile device \cite{ketal91}, photo-induced tunneling
through quantum dots has been often addressed
\cite{ketal94,betal9y,oetal97,oetal98}.

In this paper, we present physically striking results from evaluating
the general theory developed earlier, applied to driven non-linear
transport through a tunnel barrier in a one dimensional (1D) system of
interacting electrons. We report a novel frequency-locking effect,
which is signature of the systems coherent, {\em strongly} correlated
electron states. It is characteristic of the finite, non-zero range of
the interaction but does not depend on the exact shape of the driving
electric field.

By starting from the assumption that the current is driven by a local
electric field of arbitrary shape, with a DC-component $E_0(x)$ and an
additional mono-chromatic component with frequency $\Omega$,
$E(x,t)=E_{0}(x)+E_{1}(x)\cos{\Omega t}$, we find an extremely rich
behaviour of the photo-induced current-voltage characteristic
$I(V_{0};V_1,\Omega)$ as a function of various parameters, like the
range and strength of the interaction and the spatial shape of the
applied field ($V_0$, $V_1$ voltages corresponding to $E_0$, $E_1$,
respectively). (i) We first re-derive for finite range interaction the
non-trivial general expression for $I(V_{0};V_1,\Omega)$ in terms of the
(non-linear) DC current-voltage characteristic
$I_{0}(V_{0})=I(V_{0};V_1=0,\Omega=0)$ \cite{swk96,sck97,sk98}
\begin{eqnarray}
  \label{eq:1}
  I(V_{0};V_1,\Omega)=\sum_{n=-\infty}^{\infty}J_{n}^{2}(|z|)
  I_{0}^{\phantom{2}} \left(V_{0}+n\frac{\hbar\Omega}{e}\right).
\end{eqnarray}
It generalises the Tien-Gordon result to strongly correlated tunneling
objects and arbitrary shapes of the driving fields. (ii) We show the
existence of pronounced, cusp-like minima in the differential
conductance ${\rm d}I/{\rm d}V_{0}$ at integer ratios of
$eV_{0}/\hbar\Omega$ if the dimensionless interaction parameter $g\geq
2/3$, independent of how exactly the driving voltage drops. These
structures are due to sidebands that are induced by the non-linearity
of the DC current-voltage characteristics \cite{sck97,sk98}, and are
{\em independent} of the shape of the driving electric field. Finally,
(iii) we predict a cross-over between frequency scaling $\propto
\Omega^{-1}$ and $\propto \Omega^{-2}$ for delta function-like and
spatially constant driving fields, respectively. In the cross-over
region the frequency scaling is non-universal and {\em reflects} the
ranges of both, the electron-electron interaction and the range of
electric field.

It is well known that interaction-induced renormalisation of the
externally applied field influences the current in the frequency and
time domains \cite{sk96,fsk99,pb98}. Our results indicate that for 1D
correlated electrons, as long as the driving field is mono-chromatic,
the Tien-Gordon formula remains valid, but with a general argument of
the Bessel functions. The latter is given by the modulus of the complex
quantity
\begin{eqnarray}
  \label{eq:2}
  z=\frac{e}{\hbar\Omega}\int_{-\infty}^{\infty}{\rm d}x
  E_{1}(x)r(x,\Omega),
\end{eqnarray}
and contains the spatial shape of the AC-component of the driving
field, and the interaction range via the AC-conductivity
$\sigma(x,\Omega)$ without tunnel barrier,
$r(x,\Omega) \equiv \sigma(x,\Omega)/\sigma(0,\Omega)$.

In order to study the effect of finite range interactions and a
space-dependent electric field on non-linear AC-transport, we use the
1D Luttinger liquid model \cite{h81a,h81b,v95}. The Hamiltonian for a
spin-less Luttinger liquid with an impurity and subject to a
time-dependent electric field is $H = H_0 + H_{\rm t} + H_{\rm ac}$,
where $H_0$ is the Hamiltonian describing charge density excitations
with a dispersion $\omega(k)=v_{\rm F} |k|[1 + \hat{V}_{\rm ee} (k)/
\hbar \pi v_{\rm F}]^{1/2}$ ($v_{\rm F}$ Fermi velocity). It reflects
the Fourier transform of the interaction potential \cite{csk98}. We
assume a 3D screened Coulomb interaction of range $\alpha^{-1}$
projected onto a quantum wire of the diameter $d$. The dimensionless
interaction parameter of the model is $g=[1+{\hat{V}_{\rm
    ee}(k=0)}/{\hbar\pi v_{\rm F}}]^{-1/2}$.  For
    $\alpha^{-1} {\phantom{b}^{\phantom{b}_{\lesssim}}}d$,
the interaction decays exponentially and one obtains the ``Luttinger
limit'', $V_{\rm ee}(x)\propto\alpha {\rm e}^{- \alpha |x|}$
\cite{csk98}.

The tunneling barrier of height $U_{\rm t}$ is assumed to be localised
at $x=0$ \cite{kf92},
\begin{eqnarray}
\label{eq:4}
H_{\rm t}=U_{\rm t}\cos\left[2\sqrt{\pi}\vartheta (x=0)\right]
\end{eqnarray}
with the phase variable $\vartheta (x)$ giving the charge density
fluctuations in the limit of long wave-lengths $\rho(x) = k_{\rm
  F}/\pi+\partial_x \vartheta(x)/\sqrt{\pi}$.  The coupling to the
driving field is
\begin{eqnarray}
  \label{eq:6}
  H_{\rm ac}=e \int_{-\infty}^\infty {\rm d}x \rho(x) V(x,t).  
\end{eqnarray}
The voltage $V(x,t)$ is related to the field by $E(x,t)=-\partial_x
V(x,t)$. For the space-dependence of the electric field we assume
$E_{1}(x)=E_1 {\rm e}^{-|x|/a}$, with the voltage $V_1\equiv
-\int_{-\infty}^{\infty} {\rm d}xE_1(x)=-2E_1a$. When the range of the
field tends to zero, $E_1(x)$ reduces to a $\delta$-function. The
spatial dependence of the DC part of the electric field does not need
to be specified, as only the voltage $V_0= -\int_{-\infty}^\infty
{\rm d}xE_{0}(x)$ is of importance in DC transport \cite{sk96}.

The current at the barrier is given by the expectation value $I(x=0,t)
= \left\langle j(x=0,t)\right\rangle$, where the current operator is
defined via the continuity equation, $\partial_{x} j(x,t) = - e
\partial_t \rho(x,t)$. For a high barrier, the tunneling contribution
to the current can be expressed in terms of forward and backward
scattering rates which are proportional to the tunneling probability
$\Delta^2$.  The latter may be obtained in terms of the barrier height
$U_{\rm t}$ by using the instanton approximation \cite{w93}. The
result can be written in terms of the one-electron propagator $S+{\rm
  i}R$ \cite{fsk99},
\begin{equation}
\label{eq:7}
I(x=0,t) = e \Delta^2 \int_0^\infty {\rm d}\tau \, {\rm e}^{-S(\tau)}
        \sin R(\tau) \sin\left[ 
        \frac{e}{\hbar}\int_{t-\tau}^t{\rm d}t' V_{\rm eff}(t')
        \right],  
\end{equation}
with
\begin{equation}
\label{eq:8}
S(\tau) + {\rm i} R(\tau) = \frac {e^2}{\pi \hbar} 
        \int_0^{\omega_{\rm max}} \frac {{\rm d} \omega}{\omega} 
        {\cal R}e \left\{ \sigma^{-1} (x=0,\omega)\right\}
        \left[
        (1-\cos{\omega \tau})\coth
        \frac{\beta \omega}{2} 
        + {\rm i} \sin \omega\tau
        \right],
\end{equation}
where $\beta = 1/k_{\rm B}T$, $\omega_{\rm max}$ the usual frequency
cutoff that corresponds roughly to the Fermi energy \cite{solyom}, and
the AC conductivity of the system without impurity is \cite{sk96}
\begin{eqnarray}
  \label{eq:9}
  \sigma (x,\omega) = \frac {-{\rm i}v_{\rm F}e^2 \omega}
  {\hbar \pi^2}\int_0^\infty \!\!\!\!\!\!\!\!\!\!\!- {\rm d}k 
  \frac {\cos kx}{\omega^2(k)-(\omega+{\rm i}0^+)^2}\,.
\end{eqnarray}
Furthermore, the effective driving voltage is related to the electric
field by \cite{sk96}
\begin{eqnarray}
  \label{eq:10}
V_{\rm eff}(t)=\int_{-\infty}^\infty{\rm d}x
   \int_{-\infty}^t{\rm d}t' E(x,t') r(x,t-t')
   =V_0 + \frac{\hbar \Omega}{e}|z|
   \cos \left(\Omega t-\varphi_z\right) ,
\end{eqnarray}
where $|z|$ and $\varphi_z$ are, respectively, modulus and argument of
$z$ (cf. Eq.~(\ref{eq:2})). The modulus $|z|$ will be the argument of
the Bessel functions in the final result, $\varphi_z$ will represent a
phase shift of the harmonics. With the above assumptions about the
shapes of the driving field and the interaction potential one obtains
\begin{equation}
  \label{eq:12}
|z|=\frac{eV_1}{\hbar
  \Omega}\frac{1}{\sqrt{1+a^2 k^2(\Omega)}} A \left(\frac{\Omega}{v_{\rm F}
\alpha}, \frac{k(\Omega)}{\alpha}, \alpha a\right),
\end{equation}
where $k(\Omega)$ is the inverse of the dispersion relation and
\begin{equation}
  \label{eq:13}
A^2\left(u,v,w \right) 
= \frac{1}{1+u^2}\left[1+v^2\frac{(u+wv)^2}{(uw+v)^2}\right].
\end{equation}
 
In the following, we concentrate on the results for the DC component
of the current which does not depend on $x$ and is directly given by
the current at the barrier, for which we only need to know $|z|$,
 \begin{eqnarray}
   \label{eq:14}
  I_{\rm dc} &=& e \Delta^2 \int_0^{\infty} {\rm d}\tau
 {\rm e}^{-S(\tau)}
 \sin R(\tau) \sin \left(\frac{e V_0 \tau}{\hbar} \right) 
 J_0\left(2|z|\sin\frac{\Omega\tau}{2} \right)\nonumber\\
 &=&\sum_{n=-\infty}^\infty J_n^2 \left(|z|\right)I_0\left(V_0+n
 \frac{\hbar\Omega}{e}\right).         
 \end{eqnarray}
 
 The important point here is that the driven DC current is completely
 given in terms of $I_0(V_0)$, the nonlinear DC current-voltage
 characteristic of the tunnel barrier,
 \begin{eqnarray}
   \label{eq:15}
   I_0 \left(V_0\right)=e\Delta^2\int_0^\infty {\rm d}\tau
 {\rm e}^{-S(\tau)} \sin R(\tau) 
 \sin \left(\frac{eV_0}{\hbar}\tau\right).
 \end{eqnarray}
 Eqs. (\ref{eq:14}), (\ref{eq:15}) generalise results which have been
 obtained earlier \cite{tg63} but {\em without} interaction between
 the tunneling objects, and also for the Luttinger model with a
 zero-range interaction, together with a $\delta$-function like
 driving electric field \cite{swk96}.
 
 For $V_0$ much smaller than some cutoff-voltage $V_{\rm c}$ which is
 related to the inverse of the interaction range, $I_0 \propto
 V_0^{2/g-1}$. This recovers the result obtained earlier for
 $\delta$-function interaction and zero-range bias electric field
 \cite{kf92}. When $V_0\gg V_{\rm c}$, the current becomes linear
 \cite{sk98}. For intermediate values of $V_0$, $I_0$ exhibits a
 cross-over between the asymptotic regimes with a point of inflection
 near $V_{\rm c}$.  For zero-range interaction, $I_0 \propto
 V_0^{2/g-1}$ for any $V_0$. For the understanding of the driven
 current-voltage characteristic to be discussed below, this general
 shape of $I_0(V_0)$ for a finite-range interaction will turn out to
 be very important.
        
 \begin{figure}[t]
   \begin{center}
   \subfigure{\epsfig{file=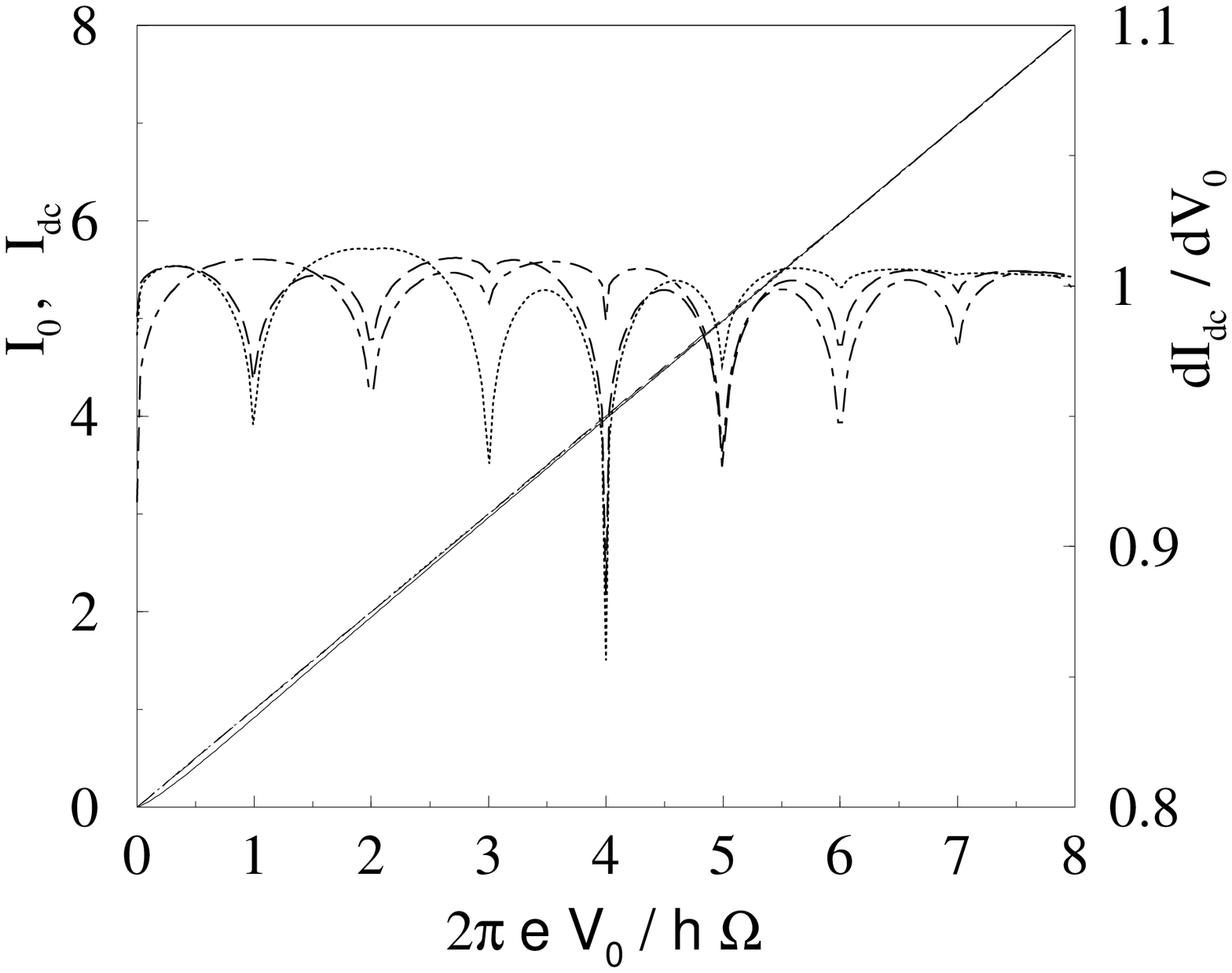, width=16pc} }
   \subfigure{\epsfig{file=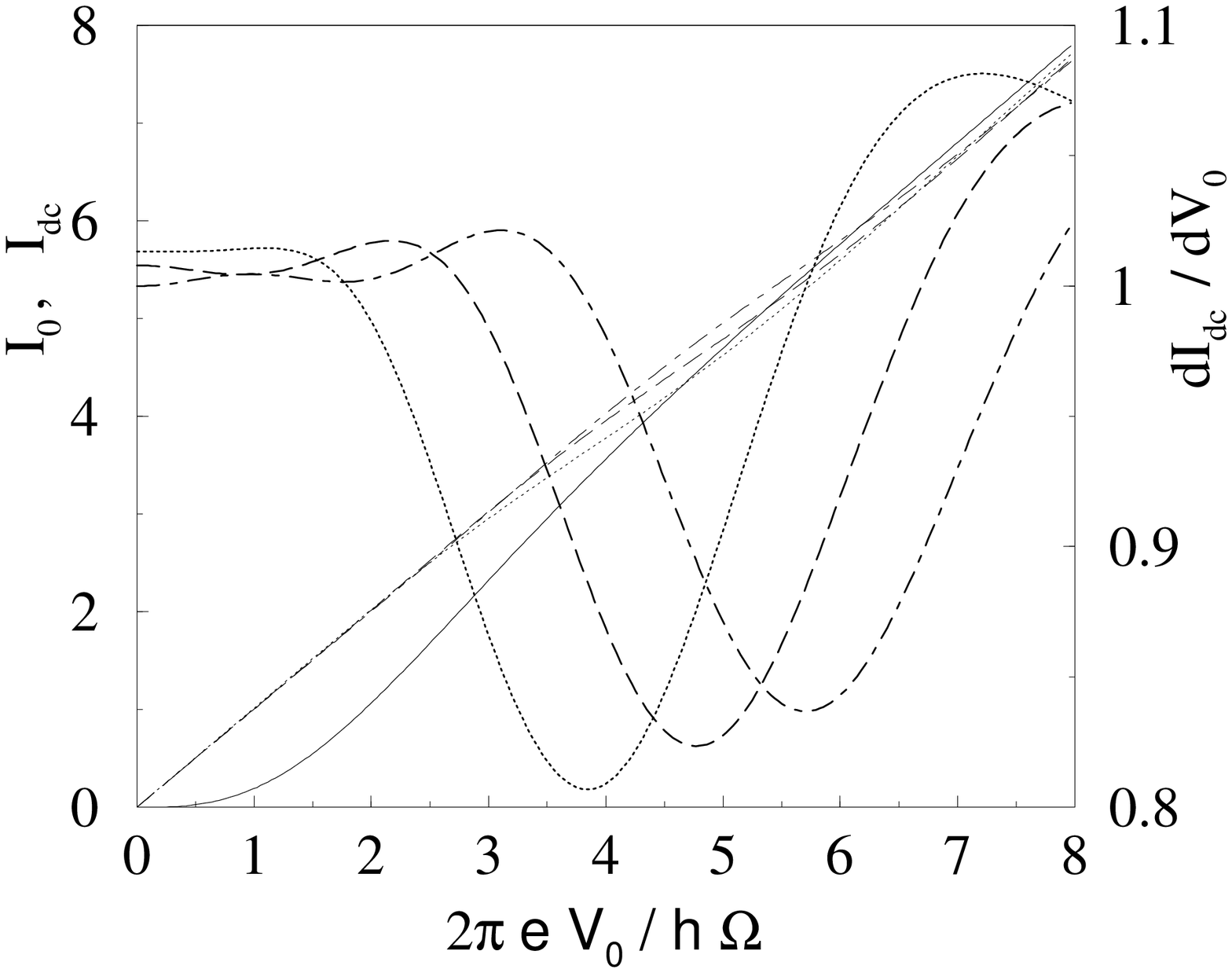, width=16pc} }
   \end{center}
     \caption[1]{Currents $I_0$, $I_{\rm dc}$ and differential 
       conductance ${\rm d}I_{\rm dc}/{\rm d}V_0$ at zero temperature
       as a function of the ratio $e V_0/\hbar \Omega$ for $g=0.9$
       (left), $g=0.5$ (right) for values $\Omega= v_{\rm F} \alpha$,
       $a=0$, and $e V_1/ \hbar v_{\rm F} \alpha =\ell$ ($\ell=5$
       dotted, $\ell=6$ dashed, $\ell=7$ dash-dotted lines). Currents
       in units of $\hbar v_{\rm F} \alpha / e R_{\rm t}$;
       differential conductance in units of $R^{-1}_{\rm t}$;
       tunneling resistance $R_{\rm t}=2 \hbar\omega^2_{\rm max}/ \pi
       e^2 \Delta^2$.}
     \label{fig:1}
 \end{figure}
 
 Figure~\ref{fig:1} shows the currents $I_0$, $I_{\rm dc}$ and the
 differential conductance ${\rm d}I_{\rm dc}/{\rm d}V_0$ as functions
 of $eV_0/\hbar\Omega$ for $g=0.9$ and $g=0.5$ for zero-range of the
 driving electric field.  For $g=0.9$ one observes sharp minima in the
 differential conductance at integer multiples of the driving
 frequency in certain regions of the driving voltage $V_1$. These can
 be understood by the following argument. When the strength of the
 interaction is not too large, the region where ${\rm d}I_{\rm
   dc}/{\rm d}V_0$ is much smaller than 1 is small compared with
 $\hbar\Omega$ such that for $eV_0\approx \hbar\Omega$, ${\rm d}I_{\rm
   dc}/{\rm d}V_0\propto
 \left(2/g-1\right)|eV_0-\hbar\Omega|^{2/g-2}$. Then, Eq.
 (\ref{eq:14}) yields near $eV_0=m\hbar\Omega$
 \begin{eqnarray}
   \label{eq:16}
   \frac{{\rm d}I_{\rm dc}}{{\rm d}V}\approx 1-J_m^2(|z|)+
   {\rm const}\cdot
   J_m^2(|z|)\left|eV_0-m\hbar\Omega\right|^{2/g-2}.
 \end{eqnarray}
 For $g>2/3$, this yields for integer $m$ the cusp-like structures
 observed in Fig.~\ref{fig:1}. For $g<2/3$, no cusps occur anymore. In
 addition, the current $I_{\rm dc}$ is depleted so strongly and over
 such a large region of the bias voltages that the regime of almost
 vanishing ${\rm d}I_{\rm dc}/{\rm d}V_0$ becomes larger than
 $\hbar\Omega$ and in general no minima near integer multiples of the
 frequency exist. As can be seen in the figure, the depths of the
 cusps depend on the driving voltage $V_1$ ($\propto |z|$) which can
 also be understood from of Eq.~(\ref{eq:16}) which shows that the
 values of the differential conductances at the voltages
 $eV_0=m\hbar\Omega$ are approximately $1-J_m^2(|z|)$.
 
\begin{figure}
\begin{center}
  \epsfig{file=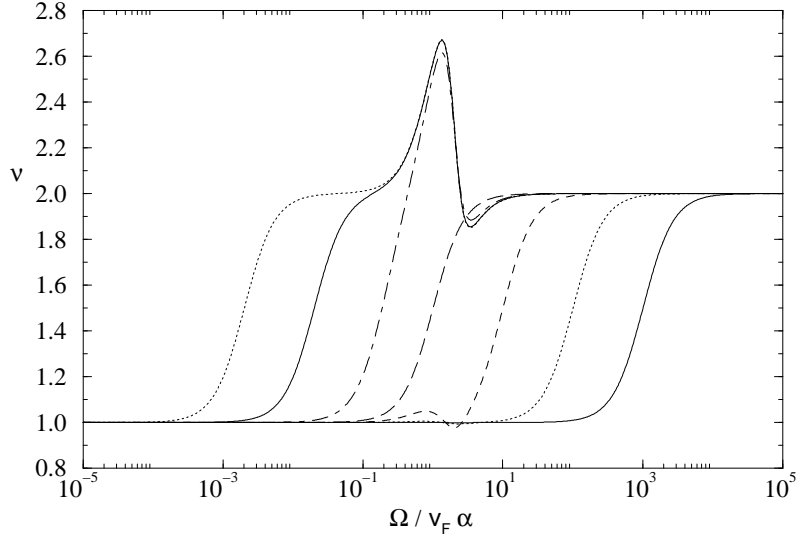, width=25pc}
\end{center}
     \caption[2]{Scaling exponent $\nu$ of the argument $|z|$ of the Bessel
       functions as a function of $\Omega$, for ranges of the driving
       field (curves from right to left) $\alpha a = 10^{-3}$,
       $10^{-2}$, $10^{-1}$, $1$, $10$, $10^{2}$, $10^{3}$, and
       $g=0.5$.}
     \label{fig:2}
 \end{figure}

 It is therefore instructive to look into the behaviour of $|z|$ as a
 function of the frequency. Figure \ref{fig:2} shows the scaling
 exponent $\nu$ determined from
 \begin{eqnarray}
   \label{eq:17}
   \nu = -v_{\rm F} \alpha \frac{{\rm d} \log |z|}
   {{\rm d} \log \Omega}.
 \end{eqnarray}
 
 We observe a non-universal cross-over between $|z|\propto
 \Omega^{-1}$, the case discussed by Tien and Gordon \cite{tg63} which
 corresponds to a driving field of zero-range ($a\to 0$), and
 $|z|\propto \Omega^{-2}$ which is obtained for a homogeneous external
 field ($a\to\infty$) \cite{w96}. Although the behaviour of $z$ depends
 strongly on the parameters of the model in the cross--over regime,
 this does not influence qualitatively the occurrence of the cusps.
 Their existence depends crucially on the finite range of the
 interaction, and the condition $g>2/3$. However, by varying $|z|$,
 the depths of the minima are changed due to the variation of
 $J_m^2(|z|)$.
 
 Given the above result for the driven DC-current, the general
 behaviour of the differential conductance as a function of
 $eV_0/\hbar\Omega$ can be straightforwardly obtained. Of special
 interest is the occurrence of cusps at $eV_0/\hbar\Omega=m$ ($m$
 integer) which appear to be quite stable against changes in the model
 parameters. A similar result has been discussed earlier \cite{lf96},
 but for a small potential barrier between fractional quantum Hall
 edge states which implies zero-range interaction. In the general case
 discussed here, the finite range of the interaction is crucial for
 obtaining the cusps, due to the absence of a linear contribution
 towards the current for small voltage which is characteristic of
 tunneling in 1D dominated by interaction. The cusps could be used to
 frequency-lock the DC part of the driving voltage.
 
 In conclusion, we have demonstrated that the result which has been
 obtained by Tien and Gordon for tunneling of non-interacting quantum
 objects in 1D driven by a mono-chromatic field localised at the
 tunnel barrier remains valid even in the presence of interactions of
 arbitrary range and shape, and for an arbitrary shape of the
 mono-chromatic driving field. The central point is that the frequency
 driven current is completely given by a linear superposition of the
 current-voltage characteristics at integer multiples of the driving
 frequency, weighted by Bessel functions.
 
 The argument of the latter contains the amplitude of the driving
 voltage only linearly but the dependence of the argument on the
 frequency and the range of the driving field is determined by its
 spatial shape. However, one can easily identify regions where the
 dependence on the frequency becomes very simple. For a driving field
 which is localised near the tunnel barrier, the integral in
 Eq.~(\ref{eq:2}) can be evaluated approximately by noting that
 $r(x,\Omega)$ varies only slowly with $x$ and can be taken out of the
 integral. Then, $|z|=e V_1/ \hbar \Omega$ which corresponds to the
 result of Tien and Gordon \cite{tg63}. In the other limit of an
 almost homogeneous electric field, $E_1=V_1/a$, one needs to
 calculate the spatial average of $r(x,\Omega)$ \cite{sk96}. This
 gives $\sigma(k=0,\Omega)/\sigma(x=0,\Omega)\approx \Omega^{-1}$,
 since $\sigma(x=0,\Omega)\approx {\rm const}$. This implies $|
 z|\propto\Omega^{-2}$. Such a frequency dependence has been
 discussed earlier for non-interacting particles \cite{w96}. Here, we
 see that it is valid under quite general assumptions also for
 interacting particles. A possible method to detect this behaviour
 experimentally is to investigate the real part of the first harmonic
 of the current through the tunnel contact and to determine the {\em
   current responsivity} which is given  by the ratio of
 the expansions of $I_{\rm dc}$ and the first harmonic to second and
 first order in $|z|$, respectively \cite{t79}.
 
 Finally, we wish to comment on the question of self-consistency
 \cite{pb98}. The driving field has been assumed {\em ad hoc} in the
 present paper, and not determined self-consistently by taking into
 account the internal interaction-induced re-arrangement of charges
 and currents as a consequence of an external field. There are
 basically two effects that have to be considered. 
 
 First, re-arrangement of the charges and currents will lead to a
 change of the spatial shape of the driving field. As discussed above,
 the present results allow to identify features that are independent
 of the spatial shape of the field and thus can be expected to survive
 even if self-consistent corrections are taken into account. Second,
 apart from the displacement current contribution, which is present
 even for $\Delta \to 0$ and renormalises $|z|$, additional
 contributions to the local field with other frequencies may be
 generated, especially in the present nonlinear case. Even the DC
 current could be changed due to mixing of harmonics present in the
 local field. If self-consistency maps contributions with different
 frequencies into the local field, possibly the above result have to
 be generalised. Strictly speaking, the Tien-Gordon form will no
 longer be valid.  However, as has been discussed earlier
 \cite{fsk99}, the amplitudes of the $n$th harmonics generated by the
 tunnel barrier when starting from a mono-chromatic external voltage
 decay extremely rapidly with $n$. Thus, when using them as a starting
 point for local field corrections, practically only the fundamental
 frequency would contribute. This would eventually lead only to
 changes in the spatial behaviour of $E_1$ and not destroy the
 existence of the cusps. In summary, we claim that local field
 corrections which need to be treated in any case beyond the mean
 field approximation are in the present context unimportant and would
 not lead to essential changes of the our findings. This concerns
 especially the existence of the cusps in the differential conductance
 at very low temperatures that are characteristic of the non-linearity
 induced by the finite-range interaction. But further, more detailed
 studies are necessary, in order to clarify the quantitative aspects
 of this issue \cite{fsk99b}.
 
 {\em Acknowledgement:} This work has been supported by the EU within
 the TMR programme, INFM via PRA(QTMD)97, Cofinanziamento (MURST)98
 and the Deutsche Forschungsgemeinschaft via the SFB 508 of the
 Universit\"at Hamburg.


\end{document}